\documentclass[aip,twocolumn,superscriptaddress,reprint,floatfix,numerical]{revtex4-1}
\usepackage{graphicx}  % needed for figures
\usepackage[breaklinks]{hyperref}
\begin{document}

\title{Random numbers from vacuum fluctuations}

\author{Yicheng Shi}
\affiliation{Department of Physics, National University of Singapore, 2
  Science Drive 3,  Singapore, 117542}
\affiliation{Center for Quantum Technologies, National University of
  Singapore, 3 Science Drive 2, Singapore, 117543}
\author{Brenda Chng}
\affiliation{Center for Quantum Technologies, National University of
  Singapore, 3 Science Drive 2, Singapore, 117543}
\author{Christian Kurtsiefer}
\affiliation{Department of Physics, National University of Singapore, 2
  Science Drive 3,  Singapore, 117542}
\affiliation{Center for Quantum Technologies, National University of
  Singapore, 3 Science Drive 2, Singapore, 117543}
\email[]{christian.kurtsiefer@gmail.com}
\date{\today}

\begin{abstract}
We implement a quantum random number generator based on a balanced
homodyne measurement of vacuum fluctuations of the electromagnetic field.
The digitized signal is directly processed with a fast randomness
extraction scheme based on a linear feedback shift register. The random bit
stream is continuously read in a computer at a rate of about 480 Mbit/s and
passes an extended test suite for random numbers.
\end{abstract}

% \pacs{}
\maketitle

\section{Introduction}
Various cryptographic schemes, classical or quantum, require high
quality and trusted random numbers for key generation and other aspects of the
protocols. 
In order to keep up with data rates in
modern communication schemes, these random numbers need to be generated at a
high rate~\cite{Gisin2002a}.
Equally, large amounts of random numbers are at the core of Monte Carlo
simulation methods~\cite{Metropolis:1987}.  
Algorithmically generated pseudo-random numbers are available at very high
rates, but are deterministic by definition and
are unsuitable for cryptographic purposes, as they may contain backdoors in
the particular algorithm used to generate them. For
applications that require unpredictable random numbers, physical random
number generators (PRNG) have been used in the past~\cite{GALTON1890} and more
recently~\cite{INTEL}. These involve measuring noisy physical processes and
conversion of the outcome into random numbers. Since it is either practically
(e.g. for thermal noise sources) or fundamentally (for certain quantum
processes) impossible to predict the outcome of such measurements, these 
physically generated random numbers are considered ``truly'' random.

Quantum random number generators (QRNG) belong to a class of physical random
number sources where the source of randomness is the fundamentally unpredictable
outcome of a quantum measurement. Early PRNG of this class were based 
 on observing the decay statistics of radioactive nuclei
 \cite{gude1985,figotin2004random}. 
 More recently, similar PRNG based on Poisson
 statistics in optical photon detection were
 implemented~\cite{Stipcevic:07,kwiat:09,Fuerst2010,Wayne:10,Wahl2011}. Different schemes use the 
randomness of a single photon scattered by a beam splitter into either of two
output ports~\cite{Jennewein2000,Stefanov2000}.
Since the reflection/transmission of the photon is  intrinsically 
random due to the quantum nature of the process, the unpredictability of the
generated numbers is ensured~\cite{Frauchiger2013}.
Other implementations of QRNGs measure the amplified spontaneous emission~\cite{Williams2010}, the
vacuum fluctuations of the electromagnetic field~\cite{Gabriel2010,Syed2011,Shen2010}, or
the intensity~\cite{Kanter2009,Sanguinetti2014} and phase noise of different light sources~\cite{Qi2010,Xu2012,Abellan2014,PhysRevLett.115.250403,Yuan2014}.

In this paper we report on a quantum random number generator based on
measuring vacuum fluctuations as the raw source of
ramdomness~\cite{Gabriel2010,Syed2011,Shen2010}. Such measurements have a very
high bandwidth compared to schemes based on photon
counting~\cite{Stipcevic:07,Fuerst2010}, and have a much simpler 
optical setup compared to phase noise measurements~\cite{Qi2010,Xu2012,Abellan2014,PhysRevLett.115.250403,Yuan2014}. Coupled with an efficient randomness extractor, we obtain an unbiased, uncorrelated stream of random bits at high speed.

\section{Implementation}
\begin{figure}
\centerline{
  \includegraphics[width=0.8\columnwidth]{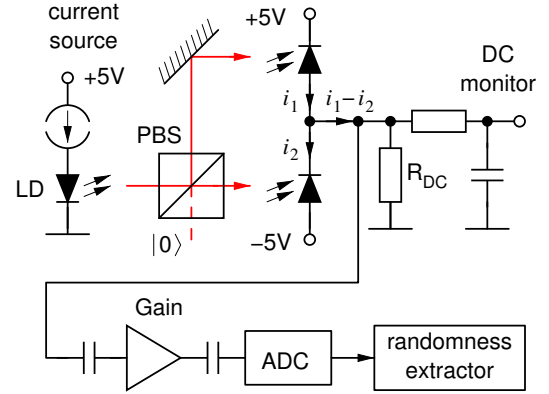}
}
\caption{Schematic of the quantum random number generator. A polarizing beam
  splitter (PBS) distributes the light of a 780\,nm laser diode equally onto two
  fast photodiodes, generating photocurrents $i_1$ and 
  $i_2$. The fluctuations in the photocurrent difference $i_1-i_2$ are
  amplified, digitized, and sent to a randomness extractor to generate
  unbiased ``true'' random numbers.}
\label{fig:schematic}
\end{figure}

Figure~\ref{fig:schematic} schematically shows the setup of our QRNG. A
continuous wave laser
(wavelength 780\,nm) is used as the local oscillator (LO) for the vacuum
fluctuations of the electromagnetic field entering the beam splitter at the
empty port. The output of the beam splitter is directed onto two pin
photodiodes, and the photocurrent difference is processed further. 
This setup is known as a balanced homodyne
detector~\cite{jakeman1974,yuen1983} and maps the the electrical field in the
second mode entering the beam splitter to the photocurrent difference
$i_1-i_2$. Here, the second input port is empty, so the homodyne measurement
is probing the vacuum state of the electromagnetic field. This field
fluctuates~\cite{Glauber:1963}, and is used as the source of randomness. As
the vacuum field is independent of external physical quantities, it can not be 
tampered with. Since the optical power impinging
on the two photodiodes is balanced, any power
fluctuation in the local oscillator will be simultaneously detected by the two
diodes, and therefore cancel in the photocurrent
difference~\cite{yuen1983,schumaker1984}. In  
an alternative view, the laser beam can be seen as generating photocurrents
$i_1,i_2$ with a shot noise power proportional to the average optical power. The
shot noise currents from the two diodes will add up because they are
uncorrelated, while amplitude fluctuations in the laser intensity (referred to
as classical noise) represented by the average current of the photodiodes does
not affect the photocurrent difference.

The power between the two output ports is balanced by rotating the laser diode
in front of a polarizing beam splitter (PBS). The output light leaving 
the PBS is detected by a pair of 
reversely biased silicon pin photodiodes (Hamamatsu S5972) connected in
series to perform the current subtraction. The balancing of the photocurrents
is monitored by observing the voltage drop across a resistor $R_D$ providing a
DC path for the current difference from the common node to ground. The
fluctuations above 20\,MHz are amplified by a transimpedance amplifier with a
calculated effective transimpedance of $R_{\rm eff}\approx540\,$k$\Omega$.

\begin{figure}
\includegraphics[width=0.9\columnwidth]{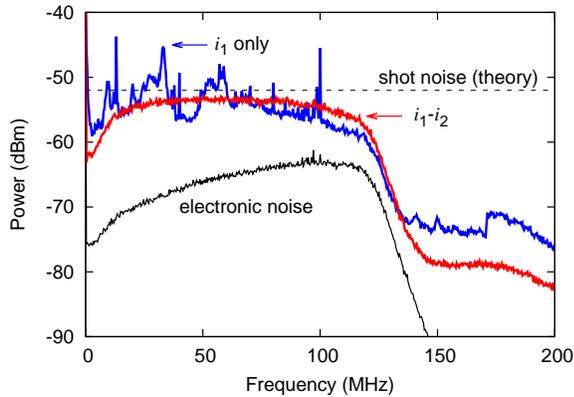}
\caption{Noise levels measured after amplification into a bandwidth
  $B=1\,$kHz. Between 20 and 120\,MHz, the total noise is measured from the
  photocurrent difference 
  $i_1-i_2$ with a balanced optical power impinging on both photodiodes and
  approaches the theoretical shot noise level of -52\,dBm (dashed trace) given by
  (\ref{eq:shot_noise}). The current $i_1$ of a single photodiode reveals
  colored classical amplitude noise. The electronic
  noise is measured without any optical input.}
\label{fig:spectrum}
\end{figure}

To ensure that the fluctuations at the output of the amplifier are
dominated by 
quantum fluctuations of the vacuum field, the spectral power density at the
output of the amplifier is measured (see
Fig.~\ref{fig:spectrum}). With an optical power of 3.1\,mW 
received by each photodiode corresponding to an average photocurrent $I=1.7$\,mA, a noise power of $P=-53.5\,$dBm (at 75\,MHz) in
a bandwidth of $B=1\,$kHz was measured. This is about 1.5\,dB lower
than the theoretically expected shot noise value (dashed trace) of 
\begin{equation}\label{eq:shot_noise}
P=\frac{4eIB{R_{\rm eff}}^2}{Z}\approx-52 \, \rm{dBm}\,,
\end{equation}
where $e$ is the electron charge and $Z=50\,\Omega$ the load impedance.
The difference is compatible with uncertainties in determining the
transimpedance of the amplifier.
The measured total noise after the amplifier has a relatively flat power density in the
range of 20 to 120\,MHz, while the high pass filters in the circuit
suppress low frequency fluctuations. The high end of the
pass band is defined by the cutoff frequency of the amplifier. To illustrate
the effectiveness of removing classical noise in the photocurrents, the
spectral power density of a photocurrent generated from a single diode is also
shown. Strong spectral peaks at various radio frequencies appear
that enter  the system probably via the laser diode current. For
completeness, the spectral power density of the electronic noise
of the amplifier is recorded without any light input, and found to be
at least 10\,dB below the total noise level, i.e., the total noise is
dominated by quantum fluctuations.

\begin{figure}
  \includegraphics[width=0.9\columnwidth]{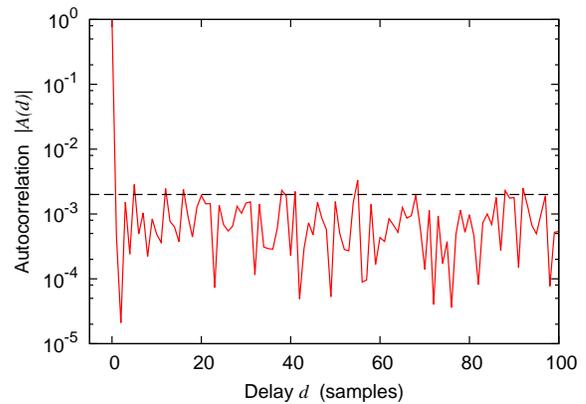}
  \caption{Autocorrelation of the total noise signal sampled at 60\,MHz,
    computed over $10^6$ samples (solid line), compared with the $2\sigma$
    confidence level (dashed line).} 
\label{fig:autocorrelation}
\end{figure}

The amplified total noise signal is digitized into signed 16 bit wide words
$x_i$ at a sampling rate of 60\,MHz with an analog to digital converter
(ADC). The sampling rate is set to be lower than 
the cut-off frequency of the noise signal in order to avoid temporal
correlation between samples. As shown in
Fig.~\ref{fig:autocorrelation}, the normalized autocorrelation
\begin{equation}
A(d)=\langle x_i\,x_{i+d}\rangle_n/\langle x_i^2\rangle_n
\end{equation}
evaluated over $n=10^6$ measured samples falls into the expected $2\sigma$
confidence interval which indicates no significant correlation between samples.

\section{Entropy estimation}
The total noise we measured before the ADC consists of both
quantum noise and the electronic noise of the detector.  To determine how much
randomness we can safely extract from the system in the sense that it
originates from a quantum process, it is necessary to quantitatively estimate
the entropy contributed by the quantum noise.

To estimate the entropy of the quantum noise $H(X_q)$, we assume that the
measured total noise signal $X_t=X_q+X_e$ is the sum of independent random
variables $X_q$ for the quantum noise, and $X_e$ for the electronic
noise.\cite{Ma2012, Sanguinetti2014}. 
Furthermore, all three
variables $X_q$, $X_e$ and $X_t$ are assumed to have discrete values
between $-2^{15}$ and $2^{15}-1$.
Since the origin of electronic noise is uncertain, we take the
worst case scenario that the adversary gains full knowledge of the electronic
noise, i.e., is able to predict the exact outcome of variable $X_e$
at any moment. In this case, the accessible amount of randomness in the acquired
total noise signal is quantified by the conditional entropy $H(X_t|X_e)$,
i.e. the amount of entropy left in the total signal, given full knowledge of
the electronic noise $X_e$. As the variables are assumed to be additive and
independent, the conditional entropy is calculated as
$H(X_t|X_e)=H(X_q+X_e|X_e)=H(X_q|X_e)=H(X_q)$.

\begin{figure}
  \includegraphics[width=0.9\columnwidth]{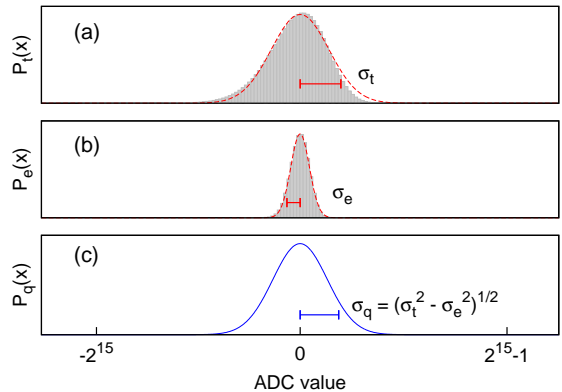}
  \caption{Probability distribution of the measured total output noise with
    variance ${{\sigma}_t}^2$ (a), electronic noise with variance
    ${{\sigma}_e}^2$ (b), and the estimated quantum noise with variance
    ${{\sigma}_q}^2$ (c). The filled areas in (a), (b) show the actual
    measurements over $10^9$ samples, the solid lines approximate the Gaussian distributions.}
  \label{fig:distribution}
\end{figure}

The variance of the total noise, $\sigma_t^2$, is given by the sum of the
variances $\sigma_q^2$ for the quantum noise, and $\sigma_e^2$ of the
electronic noise.  In an ensemble of $10^9$ samples, we find $\sigma_t=4504.41$
and $\sigma_e=1481.8$, which is measured by switching off the laser (see
Fig.~\ref{fig:distribution}).
Note that for the total noise, the observed distribution is slightly skewed
compared to a Gaussian distribution [solid line in
Fig.~\ref{fig:distribution}(a)]. We believe this is due to a distortion in the
digitizer.
Assuming the quantum noise $X_q$ has a Gaussian
distribution~\cite{Glauber:1963}, we would assign
$\sigma_q^2=\sigma_t^2-\sigma_e^2\approx 4253.7^2$.
To estimate the entropy
for a Gaussian distribution, we use the Shannon entropy
\begin{equation}\label{eq:shannon}
H(X_q)=\sum_{x=-2^{15}}^{2^{15}-1} -p_{q}(x)\log_2 p_{q}(x)\,,
\end{equation}
where $p_q(x)$is the probability distribution of the quantum noise $X_q$ with
variance 
$\sigma_q^2$. Since $\sigma_q\gg1$, $H(X_q)$ can be well approximated by 
\begin{equation} \label{eq:differential_entropy}
\int\limits_{-\infty}^{+\infty}\!\!-{f(x)}\log_2{f(x)}\,\mathrm{d}x=\log_2(\sqrt{2{\pi}e}\,\sigma_q)\,,
\end{equation}
where $f(x)$ is a Gaussian probability density function with variance
$\sigma_q^2$, and $e$ the base of the natural logarithm~\footnote{One can show
  that
  $|H(X_q)-H'(X_q)|<\log_2{(\sqrt{2\pi}\sigma_q)}/(\sqrt{2\pi}\sigma_q)\approx 0.0013$\,bit  for $\sigma_q=4108$.}.
This yields 14.1 bits of entropy per 16-bit sample.

We note that this numerical estimation of entropy only serves as an upper
bound of extractable randomness, i.e. the maximum possible amount of entropy one can
extract from the source of randomness under the assumption of a Gaussian
distribution of the independent random variables $X_q$ and $X_e$. An
alternative estimation of the entropy in $X_q$ assumes that electronic noise is
not only known to a third party, but also could be tampered
with\cite{Syed2011, Haw2014}.

\section{Randomness extraction}
In many applications, random numbers are required to be not only
unpredictable, but also uniformly distributed. As such, the raw data at the
amplifier output cannot be directly used since they are non-uniformly
distributed. Randomness extraction is the essential process required to
convert our biased raw data into a uniformly distributed binary stream at the
final output~\cite{miklos:86}. 

Various implementations of randomness extractors have been reported, such as
Trevisan's extractor and Toeplitz-hashing extractor\cite{Ma2012},
random-matrix multiplication\cite{Sanguinetti2014}, or the family of
secure hashing algorithms (SHA)\cite{Gabriel2010}. 

In this work, we use a randomness extractor based on a Linear
Feedback Shift Register (LFSR). The LFSRs are well known for quickly
generating long pseudo-random streams with little computational resources and
are in widespread use in communication applications for
spectrum whitening\cite{Krawczyk1994,Barkan2007,raey,wells2004random,4271377}.

We use a maximum length LFSR with 63 memory
cells and a two-element feedback path. Its state at any time step $t$ could be represented by 63 binary variables $s_j^t$, with a recursion relation 
\begin{eqnarray}
s_j^{t+1}&=&s_{j-1}^t\quad{\rm for}\,j=1\ldots62\label{eq:lfsrRecursion0}\,,\\
s_0^{t+1}&=&s_{62}^t\oplus s_{61}^t \label{eq:lfsrRecursion}\,,
\end{eqnarray}
where $\oplus$ denotes an exclusive-or operation. The
16 bit ADC word is serially injected into the feedback path
(\ref{eq:lfsrRecursion}) as $s_0$ with an exclusive or operation, 
\begin{equation}
s_0^{t+1}=s_{62}^t\oplus s_{61}^t\oplus d^t\,,\label{eq:lfsrRecursion2}
\end{equation}
where $d^t$ represents an input bit from the ADC word at time $t$. A reduced number of bits are extracted from $s_0$ obeying the entropy bound. To implement this efficiently in parallel for each sampled value of the
vacuum field, we add a second set of memory cells, $m_j, j=0\ldots62$, with the recursion relations
\begin{eqnarray}
m_j^{t+1}&=&s_j^t\qquad\qquad\qquad\quad\, {\rm for}\,j=0\ldots62,\label{eq:lfsrRecursion3}\\
s_j^{t+1}&=&m_j^t\oplus m_{j+1}^t\oplus d_j^t\quad {\rm for}\,j=0\ldots61,\label{eq:lfsrRecursion4}\\
s_{62}^{t+1}&=&m_{62}^t\oplus s_{0}^{t} \label{eq:lfsrRecursion5}
\end{eqnarray}
where $d_j^t$ represents the $j$-th bit of the  ADC word sampled at
$t$ for $j<16$, and $d_j^t=0$ for $j\ge16$. Recursion relations
(\ref{eq:lfsrRecursion3}-\ref{eq:lfsrRecursion5}) are equivalent to the
operation described in (\ref{eq:lfsrRecursion2}), but with all input
bits $d_j^t$ of one sampled word injected at once instead of serially. The
output bit stream is a snapshot of eight cells $m_j$ with $j=0,2,4...14$,
extracted at the ADC sampling rate (60 MHz).
The extraction ratio of 50\%  is
lower than  $14.1/16\approx88\%$ from the entropy bound estimated in (\ref{eq:differential_entropy}).
The recursion equations (\ref{eq:lfsrRecursion3}-\ref{eq:lfsrRecursion5}) and
the reduced rate extraction is implemented in a 
complex programmable logical device (CPLD, Model LC4256 from Lattice
semiconductor).

\begin{figure}
  \includegraphics[width=0.9\columnwidth]{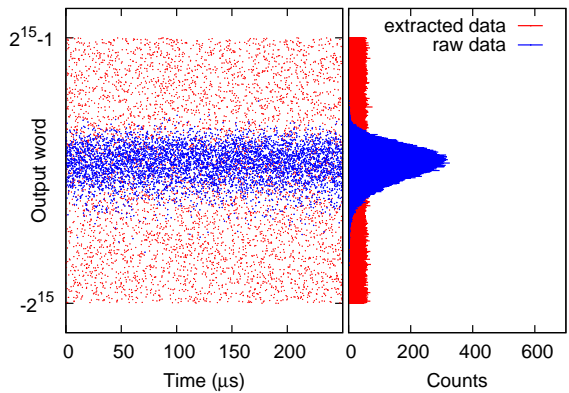}
  \caption{Distribution of random data before (blue) and after (red) the randomness extractor, shown in time domain (left) and histogram (right).}
  \label{fig:histogram}
\end{figure}

A merit of this extractor is its low circuit complexity. Unlike many secure
hashing algorithms, it can be easily implemented either in high speed or low
power technology. Therefore, the extraction process does not limit the random
number generation rate.
This scheme can receive a parallel injection of up to 63 raw bits per clock cycle while still following the extractor equations (\ref{eq:lfsrRecursion0}) and (\ref{eq:lfsrRecursion2}). With the CPLD operating at its maximum clock frequency (400 MHz), this algorithm would be able to process up to $25\times10^9$ raw input bits per second.

\section{Performance}

To evaluate the quality of the extracted random numbers, we apply two suites
of randomness tests: the statistical test suite from
NIST~\cite{NISTtestsuite}, and the ``Die-harder'' randomness test
battery~\cite{dieharder}. The output of our RNG passed both tests consistently
when evaluated over a sample of 400\,Gigabit.

Our implementation reaches an output rate of 480\,Mbit/s of uniformly
distributed random bits, with the digitizer unit sampling at 60\,MHz and
randomness extraction ratio of 50\%; this is limited by the speed limit of the
data transmission protocol we use (USB2.0). With a different transmission
protocol but the same ADC sampling, we could extract a
random bit rate of up to 60\,MHz$\times14.1$\,bits or 846\,Mbit/s. With moderate
effort, the random number generation rate can be greatly increased by extending
the bandwidth of the photodiodes, amplifiers, 
and digitizer devices, while maintaining the relatively
simple randomness extraction mechanism. Practically, the
resolution-bandwidth product of the ADC will then limit the random bit
generation rate.

\section{Conclusion}
In summary, we demonstrated a random number generation scheme by measuring
the vacuum fluctuations of the electromagnetic field. By estimating the amount
of usable entropy from quantum noise and using an efficient randomness
extractor based on linear feedback shift registers, we are able to generate
uniformly distributed random numbers at a high rate from a fundamentally
unpredictable quantum measurement.

We acknowledge the support of this work by the National Research Foundation
(partly under grant No. NRF-CRP12-2013-03) \& Ministry of Education in
Singapore, partly through the Academic Research Fund MOE2012-T3-1-009.

%\bibliographystyle{apsrev4-1}
%\bibliography{../references/reference}

\end{document}